\documentclass[12pt,a4paper,leqno,oneside]{article}
\usepackage{graphics}
\usepackage{amsthm, tabularx,hhline,longtable}
\input{epsf}
\usepackage[journal,smallskip]{mlbib}
\newtheorem{prop}{Proposition}
\newtheorem{res}{Result}
\newtheorem{simu}{S}

\theoremstyle{definition}

\theoremstyle{remark}

\begin{document}

\title{Microstructure Effects on Daily Return Volatility in Financial
Markets}
\author{Andreas Krause\thanks{University of Bath, School of Management,
Bath BA2 7AY, Great Britain, E-Mail: a.krause@bath.ac.uk}}
\date{}
\maketitle

\begin{abstract}
We simulate a series of daily returns from intraday price movements initiated
by microstructure elements. Significant evidence is found that daily returns
and daily return volatility exhibit first order autocorrelation, but trading
volume and daily return volatility are not correlated, while intraday
volatility is. We also consider GARCH effects in daily return series and show
that estimates using daily returns are biased from the influence of the level
of prices. Using daily price changes instead, we find evidence of a significant
GARCH component. These results suggest that microstructure elements have a
considerable influence on the return generating process.
\\[2mm]
\end{abstract}

It is a well known fact from a large number of empirical investigations that
financial return series exhibit specific patterns, e.g. positive
autocorrelations in volatility. Many of these patterns have been modeled using
GARCH processes as introduced by \cite*{Engle82} and \cite*{Bollerslev86} and
developed further by many other authors. \cite*{Bollerslev/Chou/Kroner92} and
\cite*{Diebold/Lopez95} give an overview of these models as well as their
empirical evidence.

Another empirical finding is the positive relation between return volatility
and trading volume. A large number of contributions address this issue with
theoretical as well as empirical investigations, see e.g.
\cite*{Foster/Viswanathan90}, \cite*{Foster/Viswanathan93a},
\cite*{Foster/Viswanathan93b}, \cite*{He/Wang95}, \cite*{Aoki99},
\cite*{Chen/Lux/Marchesi99}, \cite*{Focardi/Cincotti/Marchesi99} or
\cite*{Iori99} as few examples. A common approach in many of these models is to
assign observed effects to the processing of private information.

The model of \cite*{Clark73} allows to relate trading volume with GARCH
processes. He assumes the price change in a period of time, e.g. a trading day,
to be the sum of a large number of intraday price changes. Every trade induces
a minor price change as a new piece of independent information arrives at the
market. Therewith return volatility in a given period is proportional to the
number of trades conducted in this period. Assuming serially correlated numbers
of trades, and therewith trading volume, gives rise to GARCH effects.
\cite*{Jones/Kaul/Lipson94} show empirically that the number of trades are the
most important factor influencing return volatility, rather than trading volume
or trade sizes. A similar approach is taken by \cite*{Andersen96}.
\cite*{Lamoureux/Lastrapes90} show that with the inclusion of trading volume
GARCH coefficients are not significant.

All these models, however, do not take into account market microstructure
elements arising from nontrading periods, the bid-ask spread or inventory
control of dealers, although they are frequently considered in analyzing
intraday returns. It is well known from the literature that these elements
induce negative serial correlation of observed returns, see
\cite*{Lo/MacKinlay90} and \cite*{Roll84}. Implications for the variance of
observed returns are thus far not considered in the literature.

This paper intends to explore the effects market microstructure elements have
on daily returns and return volatility. It is beyond the scope of this paper to
give a full characterization of the observed effects by considering a large
number of parameter constellations, we concentrate on the basic properties and
leave detailed analysis for future research. We proceed as follows: the first
section introduces the model which is evaluated in section 2. Section 3
investigates GARCH effects and section 4 the behavior of a measure for intraday
volatility. Finally, section 5 concludes the findings and suggests directions
of future research.

\section{The model}

This section will introduce a very general model of price formation that
captures various elements from market microstructure. To reduce the complexity
of our model we do not consider the information flow affecting trading
behavior. Trades are only induced by investors facing a liquidity event $L_t$
in period $t$. This liquidity event is assumed to follow an AR(1) process:
\begin{eqnarray}\label{eqn1}
    L_t & = & \phi L_{t-1}+\varepsilon_t^L, \\\nonumber
    \varepsilon_t^L & \sim & iidN(0,\sigma_L^2).
\end{eqnarray}
We assume further the fundamental value of the asset in the current period to
be common knowledge and to follow a random walk:
\begin{eqnarray}\label{eqn2}
  P^*_t&=&P_{t-1}^*+\varepsilon_t^P,\\\nonumber
  \varepsilon_t^P & \sim & iidN(0,\sigma_P^2),
\end{eqnarray}
where $\varepsilon_t^L$ and $\varepsilon_t^P$ are independent. Within every
period $t$, $N_t$ trading rounds are conducted, where $N_t$ depends on the size
of the liquidity event as follows:
\begin{equation}\label{eqn3}
  N_t=Ne^{L_t},
\end{equation}
rounded to the next integer. We assume the market to be organized as a dealer
market, where dealers face inventory costs. We further suppose that only a
single dealer is present in the market conducting all trades and quoting
competitive prices. Inventory costs force the dealer after every trade to
adjust his prices. When conducting a trade at the ask, he will increase his
quote and decrease the quote when having conducted a trade at the bid, see
\cite*{Stoll78}, \cite*{Ho/Stoll80} and \cite*{Ho/Stoll81} for inventory based
models of market making.

We find the medium price $P_{t,\tau}^M$, i.e. the price in the middle between
the bid and ask price, for every trading round $\tau=1,\dots , N_t$ of period
$t$ by adjusting the fundamental value with a term denoted $\eta_{t,\tau}$
representing the inventory effect.
\begin{equation}\label{eqn4}
  P_{t,\tau}^M=P_t^*+\eta_{t,\tau}
\end{equation}
In each trading round an investor arrives at the market and decides whether to
trade at the ask, at the bid or not to trade at all. This decision depends on
the trading costs, $C_{t,\tau}^a$ and $C_{t,\tau}^b$. With $P_{t,\tau}^a$ and
$P_{t,\tau}^b$ denoting the ask and bid prices, respectively, these trading
costs are given by
\begin{eqnarray}\label{eqn5}
  C_{t,\tau}^a&=&P_{t,\tau}^a-P_t^*,\\\nonumber
  C_{t,\tau}^b&=&P_t^*-P_{t,\tau}^b.
\end{eqnarray}
We suppose these costs to transform into probabilities of trading using a logit
transformation:
\begin{eqnarray}\label{eqn6}
  \lambda_{t,\tau}^a&=&\frac{1}{1+e^{C_{t,\tau}^a}},\\\nonumber
  \lambda_{t,\tau}^b&=&\frac{1}{1+e^{C_{t,\tau}^b}}.
\end{eqnarray}
As $P_{t,\tau}^a\geq P_{t,\tau}^b$ it is easy to demonstrate that
$\lambda_{t,\tau}^a+\lambda_{t,\tau}^b\leq 1$, hence with probability
$1-\lambda_{t,\tau}^a-\lambda_{t,\tau}^b$ no trade occurs in a trading round.
Define
\begin{equation}\label{eqn7}
  I_{t,\tau}=\left\{\begin{array}{ll}
    1 & \textnormal{with probability } \lambda_{t,\tau}^a \\
    0 & \textnormal{with probability } 1-\lambda_{t,\tau}^a-\lambda_{t,\tau}^b \\
    -1 & \textnormal{with probability } \lambda_{t,\tau}^b \
  \end{array}\right..
\end{equation}
The trade size $\nu_{t,\tau}$ is assumed to be log-normal distributed and
independent of $\varepsilon_t^L$ and $\varepsilon_t^P$:
\begin{equation}\label{eqn8}
  \ln \nu_{t,\tau}\sim iidN(0,\sigma_{\nu}^2).
\end{equation}
This trade size is supposed to effect the inventory adjustment $\eta_{t,\tau}$
linearly with a scaling factor $\alpha\geq 0$. Hence we find the dynamics of
inventory adjustments as
\begin{equation}\label{eqn9}
  \eta_{t,\tau}=\eta_{t,\tau-1}+I_{t,\tau-1}\alpha\nu_{t,\tau-1}.
\end{equation}
The gross trading volume of period $t$ is given by
\begin{equation}\label{eqn10}
  V_t=\sum_{\tau=1}^{N_t}I_{t,\tau}^2\nu_{t,\tau}.
\end{equation}
Let us further assume that the dealer cannot offset his inventory between
trading periods, hence we find
\begin{equation}\label{eqn11}
  \eta_{t+1,1}=\eta_{t,N_t+1}.
\end{equation}
With $s$ denoting the constant spread applied by the dealer, the transaction
price is given by
\begin{equation}\label{eqn12}
  P_{t,\tau}=\left\{ \begin{array}{ll}
    P_{t,\tau}^M+\frac{1}{2}I_{t,\tau}s & \textnormal{if } I_{t,\tau}\not=0 \\
    P_{t,\tau-1} & \textnormal{if } I_{t,\tau}=0
  \end{array}\right..
\end{equation}
We define the final transaction price of a period as the daily price:
\begin{equation}\label{eqn13}
  P_t=P_{t,N_t}.
\end{equation}
Therewith the relevant microstructure elements, spread, inventory control and
nontrading periods have been incorporated into our model. We will now use this
model of price formation to simulate intraday prices and investigate
implications on volatility dynamics and its relation to trading volume from
daily prices.

\section{Numerical evaluation}

We use several parameter constellations to simulate a series of daily prices
and daily trading volumes using the model of section 1. For each parameter
constellation 6000 trading days have been simulated eliminating the first 1000
trading days to exclude any influences from the starting values $P_1^*=100$,
$L_0=0$, and $\eta_{1,1}=0$, so that 5000 trading days, corresponding to about
20 years of daily data, have been used for the analysis.

Throughout all simulations we assume $N=100$, $\sigma_{\nu}=.05$,
$\sigma_L=.1$, and $\sigma_P=.05$. In more detail the following parameter
constellations have been used:
\begin{simu}\label{S1}
We observe no market microstructure elements by setting $s=\alpha=0$ and the
liquidity event is serially uncorrelated ($\phi=0$).
\end{simu}
\begin{simu}\label{S2}
Here also no market microstructure elements are present, but the liquidity
event is serially correlated with $\phi=.75$.
\end{simu}
\begin{simu}\label{S3}
We assume the only market microstructure element to be the spread, i.e. $s=.25$
and  $\alpha=0$. The liquidity event is serially uncorrelated ($\phi=0$).
\end{simu}
\begin{simu}\label{S4}
We apply the same setting as in S\ref{S3}, but use serially correlated
liquidity events ($\phi=.75$).
\end{simu}
\begin{simu}\label{S5}
We assume the only market microstructure element to be inventory control, i.e.
$s=0$ and $\alpha=.1$. The liquidity event is serially uncorrelated ($\phi=0$).
\end{simu}
\begin{simu}\label{S6}
We apply the same setting as in S\ref{S5}, but use serially correlated
liquidity events ($\phi=.75$).
\end{simu}
\begin{simu}\label{S7}
Here we assume both microstructure elements to be present, the spread and
inventory control, i.e. $s=.25$ and $\alpha=.1$. The liquidity event is
serially uncorrelated ($\phi=0$).
\end{simu}
\begin{simu}\label{S8}
We apply the same setting as in S\ref{S7}, but use serially correlated
liquidity events ($\phi=.75$).
\end{simu}
These parameter settings are summarized for convenience in table \ref{tab1}.

\begin{table}
\caption{Parameter constellation of the simulations}\label{tab1}
\begin{center}
\footnotesize
\begin{tabular}{|c||c|c|c|}
\hline
  & $\alpha$ & $s$ & $\phi$ \\
\hhline{|=::=|=|=|}
 S\ref{S1} & 0 & 0 & 0 \\
\hline
 S\ref{S2} & 0 & 0 & .75 \\
\hline
 S\ref{S3} & 0 & .25 & 0 \\
\hline
 S\ref{S4} & 0 & .25 & .75 \\
\hline
 S\ref{S5} & .1 & 0 & 0 \\
\hline
 S\ref{S6} & .1 & 0 & .75 \\
\hline
 S\ref{S7} & .1 & .25 & 0 \\
\hline
 S\ref{S8} & .1 & .25 & .75\\
\hline \multicolumn{4}{l}{}\\ \multicolumn{4}{l}{$N=100$,
$\sigma_{\nu}=.05$}\\\multicolumn{4}{l}{$\sigma_L=.1$, $\sigma_P=.05$}\\
\end{tabular}
\end{center}
~\\[5mm]
\end{table}

From the simulated time series trading volume, daily returns and squared daily
returns as a measure of volatility are investigated. Daily returns are
calculated by the log-ratio of prices:
\begin{equation}\label{eqn15}
  r_t=\ln \frac{P_t}{P_{t-1}}.
\end{equation}
The results of the eight simulations are reported in table \ref{tab2}. Although
only the results and estimates of a specific realization are reported, a large
number of other realizations showed similar results, suggesting that our
findings are robust. We have not reported results on the trading volume as, not
surprisingly, it exhibits the same properties as the liquidity event, i.e.
follows an AR(1) process.

\begin{table}
 \caption{Simulation results}\label{tab2}
\begin{footnotesize}
~\\

These statistics are estimated from daily returns simulated by using the model
developed in section 1. Values exhibiting a $^*$ are significantly different
from zero at the 5\% significance level.\\[5mm]
\end{footnotesize}

\footnotesize
  \centering
  \begin{tabular}{|c||c||c|cc|c||c||c|}
  \cline{1-3}\cline{6-8}
 \textbf{S\ref{S1}} & $r_t$ & $r_t^2$&&&\textbf{S\ref{S2}} & $r_t$ & $r_t^2$ \\\hhline{|=::=::=|~~|=::=::=|}
 Mean & $1.26\times10^{-5}$ & $2.36\times10^{-7}~^*$&&&Mean & $5.99\times10^{-6}$ & $2.23\times10^{-7}~^*$ \\
\cline{1-3}\cline{6-8}
 Std. Dev. & $.000486^*$ & $3.35\times10^{-7}~^*$&&&Std. Dev. & $.000473^*$ &$3.23\times10^{-7}~^*$ \\
\hhline{|=::=b::b=|~~|=::=b::b=|}
 Lag & \multicolumn{2}{c|}{Autocorrelations} &&&Lag & \multicolumn{2}{c|}{Autocorrelations}\\
\cline{1-3}\cline{6-8}
 1 & -.006  & $.005$  &&&1 & $.009$  & $-.018$ \\
\cline{1-3}\cline{6-8}
 2 & -.009 &  $.008$ &&&2 & $.004$  & $.012$ \\
\cline{1-3}\cline{6-8}
 3 & .029  & -.017  &&&3 & .005  & .000 \\
\cline{1-3}\cline{6-8}
 4 & -.006  & -.011  &&&4 & -.013  & .002 \\
\cline{1-3}\cline{6-8}
 5 & -.004 & .005  &&&5 & .008  & -.011 \\
\cline{1-3}\cline{6-8}

 \multicolumn{3}{c}{}&\multicolumn{2}{c}{}&\multicolumn{3}{c}{}\\
 \multicolumn{3}{c}{}&\multicolumn{2}{c}{}&\multicolumn{3}{c}{}\\

\cline{1-3}\cline{6-8}
 \textbf{S\ref{S3}} & $r_t$ & $r_t^2$&&&\textbf{S\ref{S4}} & $r_t$ & $r_t^2$ \\\hhline{|=::=::=|~~|=::=::=|}
 Mean & $-2.23\times10^{-6}$ & $1.31\times10^{-5}~^*$&&&Mean & $-4.49\times10^{-6}$ & $1.22\times10^{-5}~^*$ \\
\cline{1-3}\cline{6-8}
 Std. Dev. &$.003623^*$ & $1.35\times10^{-5}~^*$&&&Std. Dev. & $.003496^*$ & $1.26\times10^{-5}~^*$ \\
\hhline{|=::=b::b=|~~|=::=b::b=|}
 Lag & \multicolumn{2}{c|}{Autocorrelations} &&&Lag & \multicolumn{2}{c|}{Autocorrelations}\\
\cline{1-3}\cline{6-8}
 1 & $-.476^*$ &  -.014 &&&1 & $-.491^*$ &  .003 \\
\cline{1-3}\cline{6-8}
 2 & -.012 &  -.007  &&&2 & .008  & .015 \\
\cline{1-3}\cline{6-8}
 3 & -.005 &  -.003  &&&3 & -.017  & -.001 \\
\cline{1-3}\cline{6-8}
 4 & .003 &  -.006  &&&4 & .011  & .004 \\
\cline{1-3}\cline{6-8}
 5 & .003  & -.009  &&&5 & .001  & -.004 \\
\cline{1-3}\cline{6-8}

\multicolumn{3}{c}{}&\multicolumn{2}{c}{}&\multicolumn{3}{c}{}\\
\multicolumn{3}{c}{}&\multicolumn{2}{c}{}&\multicolumn{3}{c}{}\\

\cline{1-3}\cline{6-8}
 \textbf{S\ref{S5}} & $r_t$ & $r_t^2$&&&\textbf{S\ref{S6}} & $r_t$ & $r_t^2$ \\\hhline{|=::=::=|~~|=::=::=|}
 Mean & $-1.82\times10^{-6}$ &$2.01\times10^{-5}~^*$&&&Mean &$8.12\times10^{-7}$ & $2.20\times10^{-5}~^*$ \\
\cline{1-3}\cline{6-8}
 Std. Dev. & $.004481^*$ & $2.87\times10^{-5}~^*$&&&Std. Dev. & $.004687^*$ &$3.06\times10^{-5}~^*$ \\\hhline{|=::=b::b=|~~|=::=b::b=|}
 Lag & \multicolumn{2}{c|}{Autocorrelations} &&&Lag & \multicolumn{2}{c|}{Autocorrelations}\\
\cline{1-3}\cline{6-8}
 1 & $-.505^*$ &  $.273^*$ &&&1 & $-.486^*$  & $.250^*$ \\
\cline{1-3}\cline{6-8}
 2 & .023 &  .009  &&&2 & -.010  & .015 \\
\cline{1-3}\cline{6-8}
 3 & -.005 &  .020 &&&3 & -.012  & -.012 \\
\cline{1-3}\cline{6-8}
 4 & -.002 &  -.009  &&&4 & .033  & -.009 \\
\cline{1-3}\cline{6-8}
 5 & -.014 &  -.008  &&&5 & -.026 & .016 \\
\cline{1-3}\cline{6-8}

 \multicolumn{3}{c}{}&\multicolumn{2}{c}{}&\multicolumn{3}{c}{}\\
 \multicolumn{3}{c}{}&\multicolumn{2}{c}{}&\multicolumn{3}{c}{}\\

\cline{1-3}\cline{6-8}
 \textbf{S\ref{S7}} & $r_t$ & $r_t^2$&&&\textbf{S\ref{S8}} & $r_t$ & $r_t^2$ \\ \hhline{|=::=::=|~~|=::=::=|}
 Mean & $-1.25\times10^{-5}$ & $2.59\times10^{-5}~^*$&&&Mean &$-4.48\times^{-6}$ & $2.67\times10^{-5}~^*$ \\
\cline{1-3}\cline{6-8}
 Std. Dev. & $.005093^*$ &$3.64\times10^{-5}~^*$&&&Std. Dev. & $.005172^*$ & $3.68\times10^{-5}~^*$ \\
\hhline{|=::=b::b=|~~|=::=b::b=|}
 Lag & \multicolumn{2}{c|}{Autocorrelations}&&&Lag & \multicolumn{2}{c|}{Autocorrelations}\\
\cline{1-3}\cline{6-8}
 1 & $-.495^*$  & $.236^*$  &&&1 & $-.501^*$  & $.242^*$ \\
\cline{1-3}\cline{6-8}
 2 & .010 &  -.017  &&&2 & .009  & .013 \\
\cline{1-3}\cline{6-8}
 3 & -.031  & -.021  &&&3 & -.012  & -.009 \\
\cline{1-3}\cline{6-8}
 4 & .033  & -.008  &&&4 & .027  & -.008 \\
\cline{1-3}\cline{6-8}
 5 & -.006  & -.011  &&&5 & -.022  & -.007 \\
\cline{1-3}\cline{6-8}
\end{tabular}
~\\[5mm]
\end{table}

Let us at first notice that correlated liquidity events, and therewith
correlated trading volume, do not affect the results significantly. For this
reason we will for the remainder of this section only consider returns arising
from uncorrelated liquidity events.

\begin{res}
  Serial correlation in trading volume does not affect the properties of
  daily return series arising from the presence of microstructure elements.
\end{res}

We further observe an increase in daily return volatility with presence of
microstructure elements. This volatility increases the more elements are added.
The equality of volatility between S1, S3, S5 and S7 can be rejected at the 1\%
significance level.

\begin{res}
  Daily return volatility increases in presence of microstructure
  elements.
\end{res}

The result derived by \cite*{Roll84} that the presence of a spread induces a
negative first order serial correlation in returns of -.5 for subsequent trades
is confirmed from our simulations also for daily returns with intraday trading.
Furthermore we do not make the assumption that trades at the bid and ask both
have a probability of .5, when neglecting trading rounds in which no trades
occur, this is only true on average. The reason for this result is that the
final trade of the day can either be at the bid or the ask. A final trade one
day at the bid, the next day at the ask, or vice versa, induces negative serial
correlation of daily returns.

We find a similar result arising from inventory control. Suppose that to the
end of a trading period a large inventory has been accumulated that could not
be offset before the last trading round, hence the price is low and the return
negative. The dealer begins the next trading period with a large inventory,
which he now tries to reduce during the trading period, causing the price to
rise the more inventory reduces and the return is positive. We therefore find
negative first order autocorrelation in daily returns. Higher order
autocorrelations are unlikely to be observed as the large number of trading
rounds within each trading period makes it unlikely that inventory has to be
reduced over several trading days. However, we can expect to find higher order
autocorrelations for less frequently traded assets, i.e. assets with a small
$N$.

The autocorrelation structure of daily returns suggests that they follow a
MA(1) process, estimates of the coefficients for this process are given in
table \ref{tab3}. Only estimates for serially uncorrelated liquidity events are
reported, estimates from the simulations using serially correlated liquidity
events show the same results. We find significant first order coefficients only
in the presence of microstructure elements. The residuals of this regression
are not serially correlated as the Durbin-Watson statistic  as well as the
Breusch-Godfrey test (both not reported here) suggest. Including higher order
moving average coefficients or autoregressive elements gives us no significant
new coefficients, but a poorer goodness of fit, so that we can confirm daily
returns to follow a MA(1) process.

\begin{res}
  Microstructure elements induce negative first order serial correlation of
  daily returns, which follow a MA(1) process.
\end{res}

\begin{table}
\caption{Estimates of MA(1) coefficients for daily returns}\label{tab3}

~~\\

\begin{footnotesize}
This table shows the least squares estimates of the MA(1) coefficients for
daily returns, $r_t=\alpha_0+\alpha_1\varepsilon_{t-1}+\varepsilon_t$. The
t-values are denoted below their estimates in parenthesis. Those estimates
significant at a 5\% level are market with a $^*$.\\[5mm]
\end{footnotesize}

\footnotesize
 \centering
\begin{tabular}{|c||c|c||c|}
\hline
  & $\alpha_0$ & $\alpha_1$ & $R^2$ \\ \hhline{|=::=|=::=|}
 S1 & $1.26\times10^{-5}$ & -.0062 & .000 \\
  & (1.8416) & (-.3985) &  \\
\hline
 S3 & $-3.01\times10^{-6}$ & $-.8042^*$ & .3837 \\
  & (-.3813) & (-95.614) &   \\
\hline
 S5 & $-1.40\times10^{-6}$ & $-.8540^*$ & .4208 \\
  & (-.1992) & (-116.006) &  \\
\hline
 S7 & $-1.21\times10^{-5}$ & $-.871^*$ & .4304 \\
  & (-1.725) & (-125.213) &  \\
\hline
\end{tabular}
~\\[5mm]
\end{table}

A final property which can be observed from table \ref{tab2} is that inventory
induces positive first order serial correlation of squared returns, i.e.
volatility. The reason for this finding is the same as for the negative first
order autocorrelation of daily returns. A high return is in most cases
associated with a large change in inventory holdings, the next trading day the
dealer offers incentives for offsetting orders to arrive at the market, hence
we will likely also find a high return, although of a different sign, causing
the volatility to be high again. Through the large number of intraday trades
this inventory offsetting is likely to be completed after one trading day, for
which reason we find no significant evidence of higher order autocorrelations.

\begin{res}\label{res5}
  Inventory control causes positive first order serial correlation of daily return volatility.
\end{res}

Throughout all parameter combinations we find no significant evidence that
daily return volatility and trading volume are correlated. We therefore have
not reported these crosscorrelations here.

\begin{res}
  Microstructure elements do not cause correlations between daily return
  volatility and trading volume.
\end{res}

We can summarize our results in the following proposition:

\begin{prop}
  The presence of microstructure elements gives rise to negative first order
  serial correlation in daily returns and positive first order correlation in
  daily return volatility, but not of any correlation between daily return
  volatility and trading volume. These properties are not affected by serial
  correlated trading volumes.
\end{prop}

\section{GARCH effects}

In nearly all financial return series evidence of GARCH effects have been
reported. We therefore have estimated the GARCH(1,1) model for the return
series generated from our simulations. The results are reported in table
\ref{tab4}.

\begin{table}
  \caption{Estimates of GARCH(1,1) coefficients for daily returns}\label{tab4}

  ~~\\
\begin{footnotesize}
  This table shows the maximum likelihood estimates of the GARCH(1,1) model for
  daily returns:\\

  $\begin{array}{l}
    r_t=\alpha_0+\alpha_1\varepsilon_{t-1}+\varepsilon_t \\
    \varepsilon_t\sim N(0,h_t) \\
    h_t=\gamma_0+\gamma_1\varepsilon_{t-1}^2+\gamma_2h_{t-1} \
  \end{array}$\\[5mm]
  The associated z-values are shown below their estimates in parenthesis and
  estimates significant at the 5\% level are marked with a $^*$.\\[5mm]
\end{footnotesize}

  \footnotesize
  \centering

  \begin{tabular}{|c||c|c||c|c|c||c|}
\hline
  & $\alpha_0$ & $\alpha_1$ & $\gamma_0$ & $\gamma_1$ & $\gamma_2$ & $R^2$ \\
\hhline{|=::=|=::=|=|=::=|}
 S1 & $1.26\times10^{-5}$ & .005 & $7.39\times10^{-8}$ & $.150^*$ & $.600^*$  & -.0001 \\
  & (1.001) & (.169) & (1.116) & (3.074) & (3.878) &    \\
\hline
 S3 & $7.05\times10^{-6}$ & $-.611^*$ & $4.86\times10^{-6}$ & $-.183^*$ & $.613^*$  & .3481 \\
  & (.611) & (-42.171) & (4.813) & (-6.801) & (5.459)   &  \\
\hline
 S5 & $2.58\times10^{-6}$ & .005 & $5.59\times10^{-6}~^*$ & $.150^*$ & $.600^*$  & -.0051 \\
  & (.043) & (.287) & (4.432) & (7.557) & (10.561) &    \\
\hline
 S7 & $-1.25\times10^{-5}$ & $-.877^*$ & $2.85\times10^{-6}$ & -.001 & $.818^*$  & .4307 \\
  & (-1.905) & (-137.602) & (.798) & (-1.093) & (3.408)   &  \\
\hline
\end{tabular}
~\\[5mm]
\end{table}

Interestingly, we also find significant coefficients when no microstructure
elements are present and hence daily returns are determined by changes in the
fundamental value, which are iid distributed. When rewriting the definition of
the return in conventional form,
\begin{equation}\label{eqna}
  \widetilde{r}_t=\ln \frac{\widetilde{P}_t^*}{\widetilde{P}_{t-1}^*}\approx
  \frac{\widetilde{P}_t^*-\widetilde{P}_{t-1}^*}{\widetilde{P}_{t-1}}=\frac{\widetilde{\varepsilon}_t}{\widetilde{P}_{t-1}^*},
\end{equation}
we see that the return is also influenced by the level of prices,
$\widetilde{P}_{t-1}^*$, which follows a MA(1) process according to
(\ref{eqn2}). The conditional variance of (\ref{eqna}) is given by
\begin{equation}\label{eqnb}
  Var\left[\widetilde{r}_t|P_{t-1}^*\right]=\left(\frac{1}{P_{t-1}^*}\right)^2\sigma_P^2.
\end{equation}
Therewith the conditional variance follows also a MA(1) process as does
$\left(\frac{1}{P^*_{t-1}}\right)^2$ and exhibits a positive first order
autocorrelation. This effect in combination with the well known difficulties in
estimating GARCH models causes the significance of the coefficients in S1 and
S2. As the same problem also arises in presence of microstructure elements, the
other estimates will also be biased as we can see from the estimates of the
mean equation. Seemingly this problem is less pronounced in presence of a
spread.

Thus far the literature has not considered the properties of conditional
variances arising from return series due to changing levels of prices. It is
beyond the scope of this paper to analyze this aspect further, instead we focus
on daily price changes, $\delta_t=P_t-P_{t-1}$,rather than returns, which we
know to be iid distributed for the fundamental value. The estimates for the
GARCH(1,1) model using daily price changes are reported in table \ref{tab6}.

\begin{table}
  \caption{Estimates of GARCH(1,1) coefficients for daily price changes}\label{tab6}

  ~~\\
\begin{footnotesize}
  This table shows the maximum likelihood estimates of the GARCH(1,1) model for
  daily price changes:\\

  $\begin{array}{l}
    \delta_t=\alpha_0+\alpha_1\varepsilon_{t-1}+\varepsilon_t \\
    \varepsilon_t\sim N(0,h_t) \\
    h_t=\gamma_0+\gamma_1\varepsilon_{t-1}^2+\gamma_2h_{t-1} \
  \end{array}$\\[5mm]
  The associated z-values are shown below their estimates in parenthesis and
  estimates significant at the 5\% level are marked with a $^*$.\\[5mm]
\end{footnotesize}

  \footnotesize
  \centering

  \begin{tabular}{|c||c|c||c|c|c||c|}
\hline
  & $\alpha_0$ & $\alpha_1$ & $\gamma_0$ & $\gamma_1$ & $\gamma_2$ & $R^2$ \\
\hhline{|=::=|=::=|=|=::=|}
 S1 & .001 & -.005 & .002 & .006 & .237 & .000 \\
  & (1.860) & (-.380) & (.346) & (.453) & (.108)   &  \\
\hline
 S3 & -.000 & $-.794^*$ & .017 & -.045 & $.831^*$  & .383 \\
   & (.154) & (-83.845) & (1.110) & (-1.783) & (4.620)   &  \\
\hline
 S5 & -.000 & $-.854^*$ & .013 & .011 & $.879^*$  & .421 \\
  & (-.191) & (-115.000) & (.920) & (1.201) & (7.037)   &  \\
\hline
 S7 & -.001 & $-.870^*$ & .017 & -.009 & $.891^*$  & .430 \\
  & (-1.905) & (-137.602) & (.798) & (-1.093) & (3.408)   &  \\
\hline
\end{tabular}
~\\[5mm]
\end{table}

We observe that the results of the mean equation are very close to those
neglecting GARCH effects as reported in table \ref{tab3}. We have not reported
these estimates for the daily price changes, but they are very close to those
of daily returns. The estimates for the ARCH component $\gamma_1$ shows no
significance for any simulation. However, with presence of microstructure
elements the GARCH component $\gamma_2$ is significant. Therewith our findings
suggest daily price changes to follow a GARCH(0,1) process with a MA(1) process
for the mean.

Although the origin of this behavior in presence of inventory control is the
positive first order autocorrelation of daily return volatility (Result
\ref{res5}), this explanation cannot be used with the spread being the only
microstructure element.

\begin{prop}
  With presence of microstructure elements daily price changes follow a
  GARCH(0,1) process.
\end{prop}

The estimates derived for our simulation are not too different from those
observed in empirical investigations. Most empirical investigations report a
relatively small ARCH component and a dominant GARCH component, which both sum
to about .9. When considering the above mentioned biases from analyzing returns
and the well known statistical problems associated with the estimation of GARCH
models, we see that at least a considerable part of the found GARCH effects may
be attributed to microstructure elements.

\section{Intraday Volatility}

We can approximate intraday volatility by comparing the highest and lowest
transaction price within a trading day, $P_t^{max}$ and $P_t^{min}$. We
therefore define
\begin{equation}\label{eqn16}
  \sigma_t^{intra}=\ln \frac{P_t^{max}}{P_t^{min}}.
\end{equation}
The statistics for $\sigma_t^{intra}$ are presented in table \ref{tab5}. We see
that without microstructure elements all transactions are conducted at the
fundamental value, which is constant throughout the day, hence the highest and
lowest price coincide. In presence of the spread as the only microstructure
element we find intraday volatility to be highly persistent over time, a unit
root cannot be rejected at any reasonable level of significance using the
Augmented Dickey-Fuller test. This behavior can be explained by the constant
absolute difference of these prices. Every transaction takes place either at
the bid or at the ask, hence the difference of the highest and lowest price is
the spread, changes in $\sigma_t^{intra}$ are only the result of changes in the
level of prices, i.e. the fundamental value, which exhibits a unit root by
construction.

\begin{table}
  \caption{Simulation results for intraday volatility}\label{tab5}

~\\[3mm]
\begin{footnotesize}
 This table shows the descriptive statistics and autocorrelations
of the intraday volatility measure $\sigma_t^{intra}$ as well as the cross
correlations with trading volume. Values being different from zero at the 5\%
significance level are marked with a $^*$.
\\[1mm]
\end{footnotesize}

  \footnotesize
  \centering
  \begin{tabular}{|c||c|c||c|c|c|c|c|}
\hline
  &  &  & \multicolumn{5}{c|}{Corr($\gamma_t$,$\gamma_{t-j}$)} \\
\hhline{|~||~|~||-|-|-|-|-|}
  & Mean & Std. Dev. & j=1 & j=2 & j=3 & j=4 & j=5 \\
\hhline{|=::=|=::=|=|=|=|=|}
 S1 & 0 & 0 & - & - & - & - & - \\
\hline
 S2 & 0 & 0 & - & - & - & - & - \\
\hline
 S3 & $.0051^*$ & $4.42\times10^{-5}~^*$ & $.998^*$ & $.996^*$ & $.994^*$ & $.992^*$ & $.990^*$ \\
\hline
 S4 & $.0049^*$ & $6.54\times10^{-5}~^*$ & $.999^*$ & $.997^*$ & $.996^*$ & $.995^*$ & $.994^*$ \\
\hline
 S5 & $.0108^*$ & $.0024^*$ & $.055^*$ & -.018 & -.010& -.024 & .012  \\
\hline
 S6 & $.0114^*$ & $.0025^*$ & $.109^*$ & .035 & .026 & .002 & .023 \\
\hline
 S7 & $.0135^*$ & $.0023^*$ & $.062^*$ & .006 & .007 & .005 & .028 \\
\hline
 S8 & $.0135^*$ & $.0024^*$ & $.083^*$ & .046 & .039 & .020 & -.001 \\
\hline
\end{tabular}
~\\[6mm]
\begin{tabular}{|c||c|c|c|c|c|c|}
\hline
 & \multicolumn{6}{c|}{Corr($\gamma_t$,$V_{t-j}$)} \\ \hhline{|~||-|-|-|-|-|-|}
 &  j=0 & j=1 & j=2 & j=3 & j=4 & j=5  \\
\hhline{|=::=|=|=|=|=|=|}
 S1 & - & - & - & - & - & - \\
\hline
 S2 & - & - & - & - & - & - \\
\hline
 S3 & -.0043 & -.0051 & -.0047 & -.0049 & -.0054 & -.0056 \\
\hline
 S4 & .0089 & .0080 & .0077 & .0075 & .0072 & .0066 \\
\hline
 S5 & $.1338^*$ & .0040 & -.0181 & -.0286 & -.0164 & -.0079 \\
\hline
 S6 & $.2306^*$ & $.1692^*$ & $.1287^*$ & $.1151^*$ & $.1015^*$ & $.0974^*$ \\
\hline
 S7 & $.1406^*$ & -.0049 & .0123 & .0069 & .0296 & .0000 \\
\hline
 S8 & $.2460^*$ & $.1683^*$ & $.1450^*$ & $.1034^*$ & $.0681^*$ & .0404 \\
\hline

\end{tabular}
~\\[2mm]
\end{table}

Inventory control, however, causes small but significant positive first order
serial correlation of intraday volatility. The argument for this finding is the
same as for the positive serial correlation of daily return volatility,
although it is of smaller magnitude. The correlation has to be smaller, because
intraday volatility is also affected by large inventory changes reversed within
the same trading day, so that volatility changes have to be less correlated. We
also find a positive correlation between current trading volume and intraday
volatility. This can easily be explained in analogy to the model of
\cite*{Clark73}. A higher liquidity event causes a larger number of trading
rounds and therewith, on average, a larger number of trades and a higher
trading volume. The larger number of trades causes the inventory adjustment,
$\eta_{t,\tau}$, to vary more within a trading day, hence the highest and
lowest transaction prices are likely to differ more, i.e. intraday volatility
is higher. The observed correlations of higher order are only the result of
correlated trading volume, but not of any persistence in the correlation of
intraday volatility. This result gives rise to our final proposition:

\begin{prop}
  Intraday volatility and trading volume are positively correlated.
\end{prop}

\section{Conclusions}

We simulated a series of daily returns incorporating market microstructure
elements and investigated the properties of these returns. Most important we
found negative first order autocorrelation of daily returns, positive first
order autocorrelation of daily return volatility, GARCH effects and a positive
correlation between trading volume and intraday volatility, but no correlation
with daily return volatility.

Many of these results can also be observed empirically, therefore the model
awaits empirical tests to specify the influence of microstructure elements on
daily returns. Future research may focus on the causes of the observed GARCH
effects and the bias in GARCH estimates of return series arising from the
influence of the price level. The generality of the model developed here allows
to apply a large variety of parameter constellations and explore the effects
arising from microstructure elements in much more detail than has been possible
here. We then may get a better understanding of the return generating process.

Our results suggest that microstructure elements should not be ignored in
analyzing daily returns. Extensions to weekly or monthly returns are
straightforward and do not change the results as the employed model does not
restrict the length of a time period. Considering these aspects may help to
find a more appropriate model for the behavior of asset prices.

\nocite{*}
\newpage
\bibliographystyle{journal}
\bibliography{vola}

\begin{thebibliography}{} \bblsize

\bibitem[\protect\citeauthoryear{{\citenf Andersen\/}}{1996}]{Andersen96}
{\bblnf Andersen, T.} \bbllb1996\bblrb: \newblock \bbllq {Return Volatility and
  Trading Volume: An Information Flow Interpretation of Stochastic
  Volatility}\bblrq.
\newblock  In: {\bbltf Journal of Finance}, {\bblvf 51}, 169--204.

\bibitem[\protect\citeauthoryear{{\citenf Aoki\/}}{1999}]{Aoki99}
{\bblnf Aoki, M.} \bbllb1999\bblrb: \newblock \bbllq {Analysis of an Open Model
  of Share Markets with Several Types of Participants}\bblrq.
\newblock In: {\bbltf Proceedings of the 4th Workshop on Economics with
  Heterogeneous Interacting Agents}. University of Genoa, Italy.

\bibitem[\protect\citeauthoryear{{\citenf Bollerslev\/}}{1986}]{Bollerslev86}
{\bblnf Bollerslev, T.} \bbllb1986\bblrb: \newblock \bbllq {Generalized
  Autoregressive Conditional Heteroskedasticity}\bblrq.
\newblock  In: {\bbltf Journal of Econometrics}, {\bblvf 31}, 307--327.

\bibitem[\protect\citeauthoryear{{\citenf Bollerslev
  et~al.\/}}{1992}]{Bollerslev/Chou/Kroner92}
{\bblnf Bollerslev, T., Chou, R.{\rm \bbland}K.Kroner} \bbllb1992\bblrb:
  \newblock \bbllq {ARCH Modelling in Finance: A review of the Theory and
  Empirical Evidence}\bblrq.
\newblock  In: {\bbltf Journal of Econometrics}, {\bblvf 52}, 5--59.

\bibitem[\protect\citeauthoryear{{\citenf Breusch\/}}{1978}]{Breusch78}
{\bblnf Breusch, T.} \bbllb1978\bblrb: \newblock \bbllq {Testing for
  Autocorrelation in Dynamic Linear Models}\bblrq.
\newblock  In: {\bbltf Australian Economic Papers}, {\bblvf 17}, 334--355.

\bibitem[\protect\citeauthoryear{{\citenf Campbell
  et~al.\/}}{1997}]{Campbell/Lo/MacKinlay97}
{\bblnf Campbell, J., Lo, A.{\rm \bbland}MacKinlay, C.} \bbllb1997\bblrb:
  \newblock {\bbltf {The Econometrics of Financial Markets}}.
\newblock Princeton: Princeton University Press.

\bibitem[\protect\citeauthoryear{{\citenf Chen
  et~al.\/}}{1999}]{Chen/Lux/Marchesi99}
{\bblnf Chen, S., Lux, T.{\rm \bbland}Marchesi, M.} \bbllb1999\bblrb: \newblock
  \bbllq {Testing for Nonlinearity in an Artificial Financial Market}\bblrq.
\newblock In: {\bbltf Proceedings of the 4th Workshop on Economics with
  Heterogeneous Interacting Agents}. University of Genoa, Italy.

\bibitem[\protect\citeauthoryear{{\citenf Clark\/}}{1973}]{Clark73}
{\bblnf Clark, P.} \bbllb1973\bblrb: \newblock \bbllq {A Subordinated
  Stochastic Process Model with Finite Variance for Speculative Prices}\bblrq.
\newblock  In: {\bbltf Econometrica}, {\bblvf 41}, 131--156.

\bibitem[\protect\citeauthoryear{{\citenf
  Diebold{\bbland}Lopez\/}}{1995}]{Diebold/Lopez95}
{\bblnf Diebold, F.{\rm \bbland}Lopez, J.} \bbllb1995\bblrb: \newblock \bbllq
  {Modeling Volatility Dynamics}\bblrq.
\newblock Federal Reserve Bank of New York Research Paper 9522.

\bibitem[\protect\citeauthoryear{{\citenf Engle\/}}{1982}]{Engle82}
{\bblnf Engle, R.} \bbllb1982\bblrb: \newblock \bbllq {Autoregressive
  Condistional Heteroskedasticity with Estimates of the Variance of UK
  Inflation}\bblrq.
\newblock  In: {\bbltf Econometrica}, {\bblvf 50}, 987--1008.

\bibitem[\protect\citeauthoryear{{\citenf Focardi
  et~al.\/}}{1999}]{Focardi/Cincotti/Marchesi99}
{\bblnf Focardi, S., Cincotti, S.{\rm \bbland}Marchesi, M.} \bbllb1999\bblrb:
  \newblock \bbllq {Self-organized Criticality and Large Crashes in Financial
  Markets}\bblrq.
\newblock In: {\bbltf Proceedings of the 4th Workshop on Economics with
  Heterogeneous Interacting Agents}. University of Genoa, Italy.

\bibitem[\protect\citeauthoryear{{\citenf
  Foster{\bbland}Viswanathan\/}}{1990}]{Foster/Viswanathan90}
{\bblnf Foster, D.{\rm \bbland}Viswanathan, S.} \bbllb1990\bblrb: \newblock
  \bbllq {A Theory of the Interday Variations in Volume, Variance, and Trading
  Costs in Securities Markets}\bblrq.
\newblock  In: {\bbltf Review of Financial Studies}, {\bblvf 3}, 593--624.

\bibitem[\protect\citeauthoryear{{\citenf
  Foster{\bbland}Viswanathan\/}}{1993a}]{Foster/Viswanathan93a}
{\bblnf Foster, D.{\rm \bbland}Viswanathan, S.} \bbllb1993a\bblrb: \newblock
  \bbllq {The Effect of Public Information and Competition on Trading Volume
  and Price Volatility}\bblrq.
\newblock  In: {\bbltf Review of Financial Studies}, {\bblvf 6}, 23--56.

\bibitem[\protect\citeauthoryear{{\citenf
  Foster{\bbland}Viswanathan\/}}{1993b}]{Foster/Viswanathan93b}
{\bblnf Foster, D.{\rm \bbland}Viswanathan, S.} \bbllb1993b\bblrb: \newblock
  \bbllq {Variations in Trading Volume, Return Volatility, and Trading Costs:
  Evidence on Recent Price Formation Models}\bblrq.
\newblock  In: {\bbltf Journal of Finance}, {\bblvf 48}, 187--211.

\bibitem[\protect\citeauthoryear{{\citenf Godfrey\/}}{1978}]{Godfrey78}
{\bblnf Godfrey, L.~G.} \bbllb1978\bblrb: \newblock \bbllq {Testing against
  General Autoregressive and Moving Average Error Models when the Regressors
  Include Lagged Dependent Variables}\bblrq.
\newblock  In: {\bbltf Econometrica}, {\bblvf 46}, 1293--1302.

\bibitem[\protect\citeauthoryear{{\citenf Godfrey\/}}{1988}]{Godfrey88}
{\bblnf Godfrey, L.~G.} \bbllb1988\bblrb: \newblock {\bbltf {Specification
  Tests in Econometrics: The Lagrange Multiplier Principle and other
  Approaches}}, \bblvol~16, {\bbltf Econometric Society Monographs Series}.
\newblock Cambridge, UK: Cambridge University Press.

\bibitem[\protect\citeauthoryear{{\citenf Hamilton\/}}{1994}]{Hamilton94}
{\bblnf Hamilton, J.} \bbllb1994\bblrb: \newblock {\bbltf {Time Series
  Analysis}}.
\newblock Princeton: Princeton University Press.

\bibitem[\protect\citeauthoryear{{\citenf He{\bbland}Wang\/}}{1995}]{He/Wang95}
{\bblnf He, H.{\rm \bbland}Wang, J.} \bbllb1995\bblrb: \newblock \bbllq
  {Differential Information and Dynamic Behavior of Stock Trading
  Volume}\bblrq.
\newblock  In: {\bbltf Review of Financial Studies}, {\bblvf 8}, 919--972.

\bibitem[\protect\citeauthoryear{{\citenf
  Ho{\bbland}Stoll\/}}{1980}]{Ho/Stoll80}
{\bblnf Ho, T.{\rm \bbland}Stoll, H.} \bbllb1980\bblrb: \newblock \bbllq {On
  Dealership Markets under Competition}\bblrq.
\newblock  In: {\bbltf Journal of Finance}, {\bblvf 35}, 259--267.

\bibitem[\protect\citeauthoryear{{\citenf
  Ho{\bbland}Stoll\/}}{1981}]{Ho/Stoll81}
{\bblnf Ho, T.{\rm \bbland}Stoll, H.} \bbllb1981\bblrb: \newblock \bbllq
  {Optimal Dealer Pricing under Transactions and Return Uncertainty}\bblrq.
\newblock  In: {\bbltf Journal of Financial Economics}, {\bblvf 9}, 47--73.

\bibitem[\protect\citeauthoryear{{\citenf Iori\/}}{1999}]{Iori99}
{\bblnf Iori, G.} \bbllb1999\bblrb: \newblock \bbllq {A Microsimulation of
  Traders Activity in the Stock Market: The Role of Heterogeneous Expectations,
  Agents Interaction and Information Flow}\bblrq.
\newblock In: {\bbltf Proceedings of the 4th Workshop on Economics with
  Heterogeneous Interacting Agents}. University of Genoa, Italy.

\bibitem[\protect\citeauthoryear{{\citenf Jones
  et~al.\/}}{1994}]{Jones/Kaul/Lipson94}
{\bblnf Jones, C., Kaul, G.{\rm \bbland}Lipson, M.} \bbllb1994\bblrb: \newblock
  \bbllq {Transactions, Volume, and Volatility}\bblrq.
\newblock  In: {\bbltf Review of Financial Studies}, {\bblvf 7}, 631--651.

\bibitem[\protect\citeauthoryear{{\citenf
  Lamoureux{\bbland}Lastrapes\/}}{1990}]{Lamoureux/Lastrapes90}
{\bblnf Lamoureux, C.{\rm \bbland}Lastrapes, W.} \bbllb1990\bblrb: \newblock
  \bbllq {Heteroskedasticity in Stock Return Data: Volume versus GARCH
  Effects}\bblrq.
\newblock  In: {\bbltf Journal of Finance}, {\bblvf 45}, 221--229.

\bibitem[\protect\citeauthoryear{{\citenf
  Lo{\bbland}MacKinlay\/}}{1990}]{Lo/MacKinlay90}
{\bblnf Lo, A.{\rm \bbland}MacKinlay, C.} \bbllb1990\bblrb: \newblock \bbllq
  {An Econometric Analysis of Nonsynchronous-Trading}\bblrq.
\newblock  In: {\bbltf Journal of Econometrics}, {\bblvf 45}, 181--212.

\bibitem[\protect\citeauthoryear{{\citenf Roll\/}}{1984}]{Roll84}
{\bblnf Roll, R.} \bbllb1984\bblrb: \newblock \bbllq {A Simple Implicit Measure
  of the Effective Bid-Ask Spread in an Efficient Market}\bblrq.
\newblock  In: {\bbltf Journal of Finance}, {\bblvf 39}, 1127--1140.

\bibitem[\protect\citeauthoryear{{\citenf Stoll\/}}{1978}]{Stoll78}
{\bblnf Stoll, H.} \bbllb1978\bblrb: \newblock \bbllq {The Supply of Dealer
  Services in Securities Markets}\bblrq.
\newblock  In: {\bbltf Journal of Finance}, {\bblvf 22}, 1133--1151.

\end{thebibliography}
\end{document}